\documentclass{aa}  
\usepackage{graphicx}
\usepackage{xcolor}
\usepackage{textcomp}
\usepackage{scalefnt}   
\usepackage[varg]{txfonts}

\usepackage{hyperref}
\newbox\grsign \setbox\grsign=\hbox{$>$} \newdimen\grdimen \grdimen=\ht\grsign
\newbox\simlessbox \newbox\simgreatbox
\setbox\simgreatbox=\hbox{\raise.5ex\hbox{$>$}\llap
     {\lower.5ex\hbox{$\sim$}}}\ht1=\grdimen\dp1=0pt
\setbox\simlessbox=\hbox{\raise.5ex\hbox{$<$}\llap
     {\lower.5ex\hbox{$\sim$}}}\ht2=\grdimen\dp2=0pt

\newbox\simppropto
\setbox\simppropto=\hbox{\raise.5ex\hbox{$\sim$}\llap
     {\lower.5ex\hbox{$\propto$}}}\ht2=\grdimen\dp2=0pt

\begin{document}

\title{\fontsize{14.3}{10}\selectfont{Precise boron abundance in a sample of metal-poor stars from far-ultraviolet lines}}
\titlerunning{Boron abundance in metal-poor stars }
 \authorrunning{M. Spite et al.}
 
 \author{\fontsize{10}{10}\selectfont{
 M. Spite \inst{1},
 B. Barbuy \inst{2},
 K. Tan \inst{3}
}
}

\institute{
LIRA, Observatoire de Paris, Universit\'e PSL, CNRS, 5 Place Jules Janssen, 92190 Meudon, France
\and
Universidade de S\~ao Paulo,  IAG, Departamento de Astronomia, 05508-090 S\~ao Paulo, Brazil
\and 
National Astronomical Observatories, Chinese Academy of Sciences, Beijing 100101, China
}
            
\date{Received 16 July 2025; accepted 2 September 2025}
\abstract
{The light elements beryllium (Be; $Z=4$) and boron (B; $Z=5$) are mainly produced by spallation reactions between cosmic rays and carbon (C; $Z=6$), nitrogen (N; $Z=7$), and oxygen (O; $Z=8$) nuclei. Only traces of Be or B would have been produced in the Big Bang, but there could be a contribution from the $\nu$-process in   type II supernovae. Their abundances at very low metallicities have been debated in the literature, with the aim of
 understanding their origin.}
 {Our aim is to derive the boron abundance in a sample of metal-poor stars based for the first time on observations with the
 STIS spectrograph on board  the Hubble Space Telescope, using clean B lines measured in space ultraviolet.}
 {We identified a measurable line of \ion{B}{I} at 2089.6\,{\AA}.
 In our sample of metal-poor warm stars, this line is practically free from blending lines, 
 and for this reason the precision of the presently derived boron abundances is unprecedented.} 
 {We find that in the interval $\rm -2.6<[Fe/H]<-1.0$, 
the slope of the relation A(B) versus [Fe/H] is significantly larger than 1, and thus steeper than that obtained with Be abundances. 
As a consequence, we find in this interval of metallicity a B/Be ratio that slightly increases with [Fe/H]. 
 Since at $\rm [Fe/H]=-1$ the abundance of B is already close to the solar abundance, there should be a break in the B enrichment at a metallicity of about $\rm [Fe/H]=-1$.}
  {}
   \maketitle

\section{Introduction}
Our aim in the present paper is to better understand the behaviour of the abundances of B at low metallicities. We chose to check these abundances in a sample of metal-poor stars where the Li abundance is compatible with the `Spite plateau' \citep[see e.g.][]{BonifacioAARev25}. This plateau should correspond to the abundance of lithium in the gas which formed the stars, and thus where we can expect that the B and Be abundances are also witnesses of the abundance of these elements in this primordial gas.\footnote{In this paper we will not discuss the fact that the abundance of Li measured in warm very metal-poor stars, is below the value deduced from the Planck observations in the frame of the standard Big-Bang, by about 0.4\,dex following \citet{PlanckColl-XIII-16}.} 

The light elements Li, Be, and B are indeed very fragile; they are destroyed by (p, $\alpha$)
reactions as soon as the temperature reaches 2.5$\times$10$^{6}$, 3.5$\times$10$^{6}$, and 5.0$\times$10$^{6}$\,K, respectively. As a consequence, in a stellar atmosphere, if the convective zone is deep enough, these elements are brought to deeper layers, where they are little by little destroyed. A priori the primitive abundance of Li (the most fragile of these elements), can be only preserved in warm metal-poor dwarfs with an effective temperature higher than 5900\,K and $\rm -2.8<[Fe/H]<-1.8$\footnote{We adopt the classical notation that for element X: $\rm A(X) = log(N_{X} / N_{H}) + 12$ and $\rm [X/H] = log(N_{X} / N_{H})_{star} - log(N_{X} / N_{H})_{Sun}$.}
\citep[see e.g.][]{spite82b,spite82a,Bonifacio-Li-07,Hosford-Li-10,Reggiani-Li-17,MatasPinto-Li-21}.
In these stars we can expect that the primitive abundance of Be and B has also been preserved \citep{boesgaard23}.

The Big Bang nucleosynthesis \citep[e.g.][]{boesgaard85,Coc-BBN-16,Cyburt-BBN-16,pitrou-BBN-18}
produces $^{7}$Li, and Li is also produced later in novae \citep{Borisov24}. On the other hand, 
Be and B are mainly built by Galactic cosmic rays (GCRs) bombarding 
C, N, O interstellar atoms, in particular in the surroundings of supernovae; this is called the spallation process, and was first suggested by
\citet{reeves70} and \citet{meneguzzi71}. As a consequence it was first expected that Be and B should behave as secondary elements. 

For a review, \citet{tatischeff18} describe the
spallation process and production of the light element isotopes
$^{6}$Li, $^{7}$Li, $^{9}$Be, $^{10}$B, and $^{11}$B in the interstellar medium (ISM).

\citet{prantzos12} computed models to describe the galactic chemical evolution
of Be and B. He indicated that the isotope $^{10}$B, as well as the Be abundances,
are well reproduced by the spallation process.
Another process is required to explain the higher abundance of $^{11}$B, given
that $^{10}$B and $^{11}$B constitute 19.9\% and 80.1\% of the B solar composition
\citep{asplund09}.
\citet{woosley90} introduced the concept of neutrino-process, or $\nu$-process, based on the 
fact that as massive stars collapse, neutrinos are emitted in
large numbers. Part of this huge number of neutrinos interact with nuclei
and produce mainly $^{7}$Li and $^{11}$B among the light elements.
Production of $^{11}$B is due to neutrino interactions
with carbon in the carbon layer, and to a lesser extent in the helium layer.
\citet{woosley95} confirmed that
whereas essentially no $^{9}$Be is made in massive stars, 
a small amount of $^{10}$B and a much larger amount of $^{11}$B are
produced in massive stars. Therefore,
the $\nu$-process occurring in the explosion of supernovae type II, or in core-collapse
supernovae (CCSN), is a complementary probable source of $^{11}$B.
\citet{kusakabe19} gave a revisited study of the $\nu$-process in
the formation of the light elements.
\citet{yoshida08} computed the $\nu$-process using a nuclear
reaction network of 291 species, and discussed the uncertainties in the $\nu$-process yields.
\citet{nakamura10,nakamura10b} made
calculations of the $\nu$-process in intermediate mass (2--16\,$M_{\odot}$) supernovae of type Ic, using the nuclear
reaction network established by \citet{yoshida08}, and also succeeded  
in producing a significant amount of $^{11}$B.

The abundance of beryllium is rather well studied, despite difficulties with the blending of the
\ion{Be}{II} 3130\,{\AA} line and the lower flux in this near-ultraviolet region close to
the atmospheric cutoff. Observations down to metallicites $\rm [Fe/H]\sim-3.0$ are
available \citep{Boesgaard11,smiljanic21,BonifacioAARev25}, and they are compatible with a linear decrease in the Be abundance with metallicity.

Boron instead has been observed very little, for the main reason that lines are only found in the space ultraviolet. The B abundance has been measured in only ten metal-poor dwarfs with a Li abundance compatible with the plateau, and having metallicities $\rm [Fe/H]\leq-1.0$.
All the studies of the B abundance in metal-poor stars \citep{duncan97,primas98,primas99,garcia-lopez98,tan10} are based on the same Hubble Space Telescope (HST) spectra obtained in the late 1990s and retrieved from the HST Science Archive Database. The HST was equiped with the Goddard High Resolution Spectrograph (GHRS) with a resolution $R \approx 25\,000$  at $\lambda=250$\,nm and a signal-to-noise ratio $S/N>50$.
The abundance of B was deduced from the resonance line of \ion{B}{I} at 2496.7\,{\AA},  but this line is in a very crowded region and is severely blended, in particular by a rather strong \ion{Co}{I} line, and thus these B abundances cannot be considered as very well established.

\section {Observations and line data}
\label{sec:obs}

For the first time after about 25 years we could use new HST spectra.
They were obtained in a farther-ultraviolet spectral region, with the STIS spectrograph and the E230H echelle grating.
The observations are described in detail in Table 1 of \citet{peterson20}.\footnote{The E230H spectra for HD\,19445, HD\,140283 and HD\,84937 were obtained with programmes GO-14672 and GO-14161 (PI: Peterson). 
For HD\,94028, HD\,160617 and HD\,76932, we used also spectra obtained with the programmes GO-8197 and GO-9804 (PI: Duncan) and for HD\,140283, with the programme GO-7348 (PI: Edvardsson).}
The resolving power of these new spectra is $R \approx 110,000$ and the signal-to-noise $S/N>50$.
In this work these very high-resolution spectra were used to derive the B abundance from the ultraviolet resonance line at 2089.6\,{\AA}. The  lines in this region of the spectra are generally much cleaner, with fewer blends.
On this basis, we hope to improve the relation between the B abundance and the metallicity in our Galaxy.

\begin{table}
\centering
\caption[4]{\ion{B}{I} line list and oscillator strengths following the NIST data base \citep{NISTdataBase}.}
\begin{tabular}{lccll}
\hline
\noalign{\smallskip}
Multiplet& Wavelength &$\chi_{\rm ex}$&   $\log gf$ & Qual  \\
         & {\AA}      & eV            &             &       \\
\hline
$2s^{2}2p-2s2p^{2}$  &2088.889 & 0.000 &     $-$1.024   & B+   \\
                     &2089.556 & 0.002 &     $-$1.73    & B+   \\
                     &2089.570 & 0.002 &     $-$0.77    & B+   \\ 
\\
$2s^{2}2p-2s^{2}3s$  &2496.769 & 0.000 &     $-$0.80    & A    \\
                     &2497.723 &	0.002 &     $-$0.503   & A    \\ 
\hline
\label{tab:linelist}
\end{tabular}
\end{table} 

The precise wavelengths and oscillator strengths of \ion{B}{I} lines are taken from the NIST standard reference data base \citep{NISTdataBase}, and are given in Table \ref{tab:linelist}. 
For the other lines used in the computation of the synthetic spectra, the data from the 
 Vienna Atomic Lines Data Base VALD 
\footnote{https://www.astro.uu.se/}
\footnote{https://www.astro.uu.se/valdwiki/Acknowledgement} 
were adopted.

\begin{figure}
    \centering
    \includegraphics[width=9.0cm]{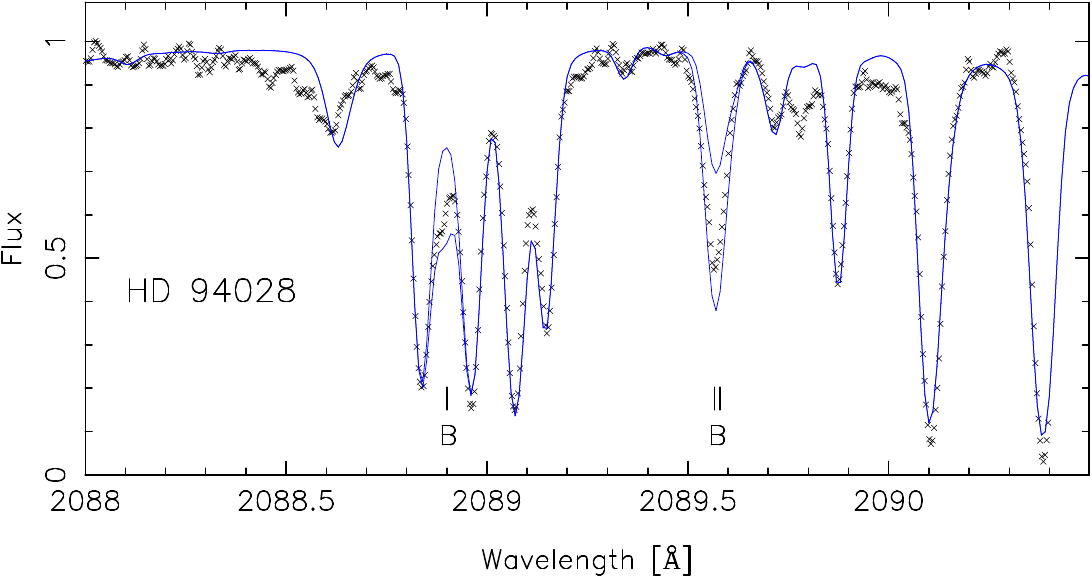}
    \includegraphics[width=9.0cm]{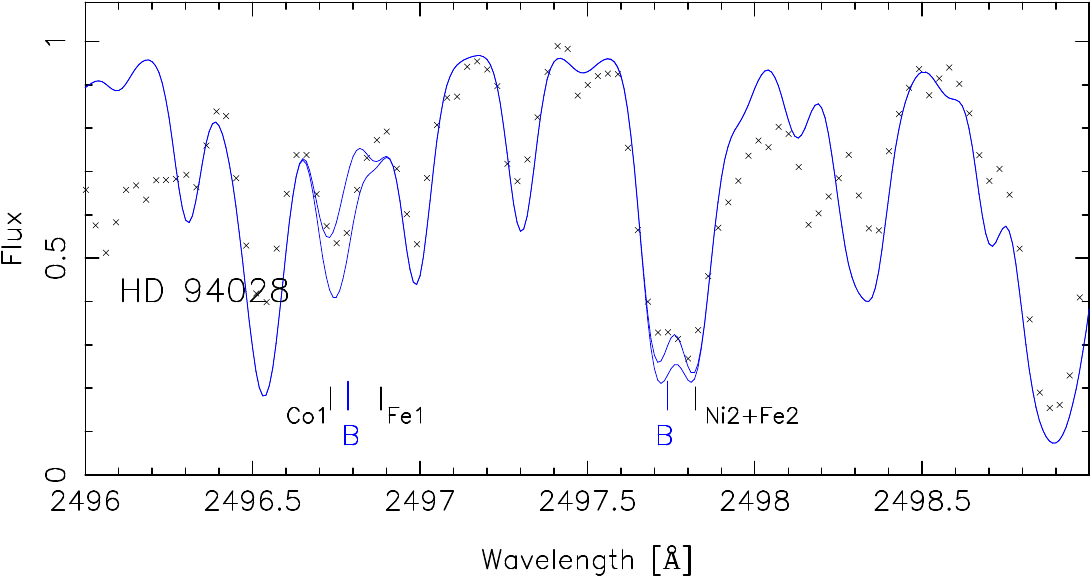}
    \caption{\ion{B}{I} 2088.889, 2089.556, and 2089.570\,{\AA} lines 
    (upper panel) and 2496.7\,{\AA} line (lower panel) in HD\,94028. 
    In both cases the figure represents 3\,{\AA} ~in the region of the B lines.
    The synthetic spectra (blue lines) were computed for A(B)=1.0 and A(B)=1.5\,.} 
    \label{fig:hd94boro}
\end{figure}

In Fig. \ref{fig:hd94boro} we show the synthetic spectra of HD\,94028 in the region of the two \ion{B}{I} multiplets of Table \ref{tab:linelist}.
The quality (S/N and resolving power) of the new STIS spectra around 2089\,{\AA} (top) is much better than the HST spectra previously used to determine the abundance of B (bottom). Moreover, it can be seen that the B lines around 2496\,{\AA} are severely blended;   the first line of the first multiplet at 2088.889\,{\AA} is also severely mixed with rather strong lines of \ion{Fe}{II} towards the blue and \ion{Ni}{I} towards the red. Therefore, for the present paper, we   measured the abundance of boron using only the line of \ion{B}{I} at 2089.6\,{\AA}. This line, the sum of two \ion{B}{I} lines at 2089.556 and 2089.570\,{\AA} is very clean, and is clearly the best indicator of the abundance of boron.
This very good B indicator has not been used before,  except for testing the ratio $\rm ^{11}B/^{10}B$~ in HD\,76932  \citep{rebull98}, mainly because of the difficulty in observing this region;  HST/STIS requires  exposure times that are around ten times longer  than in the near-ultraviolet, in part also due to the higher resolution of the STIS E230H grating with respect to that of E230M.

\begin{table}
\begin{center}   
\caption[]{Stellar parameters adopted from \citet{spite24}.}
\label{tab:param}
\begin{tabular}{lcccc}
\hline
\hbox{Star} & \hbox{$T_{\rm eff}$}  &  \hbox{$\log g$} &     [Fe/H]& \hbox{$v_{\rm t}$}\\
\hbox{}             &    K &                 &            & km/s   \\
\noalign{\vskip 0.05cm}
\noalign{\hrule}
\noalign{\vskip 0.05cm}
\hbox{HD~19445}  & 6070 &  4.40 & $-2.15$ & 1.3 \\  
\hbox{HD~76932}  & 6000 &  4.10 & $-0.94$ & 1.3 \\
\hbox{HD~84937}  & 6300 &  4.00 & $-2.25$ & 1.3 \\    
\hbox{HD~94028}  & 6050 &  4.30 & $-1.40$ & 1.2 \\  
\hbox{HD~140283} & 5750 &  3.70 & $-2.57$ & 1.4 \\  
\hbox{HD~160617} & 6000 &  3.90 & $-1.80$ & 1.2 \\ 
\hline	 
\end{tabular} 
\end{center}  
\end{table}

\section{Stellar parameters and B abundance}   \label{sec:abund}

The atmospheric parameters of the sample stars have been taken from \citet{peterson20,spite22,spite24}. They are based on Gaia photometry \citep{gaia18a,arenou18}, corrected for reddening following \citet{lallement19} and combined with PARSEC isochrones \citep{bressan12,marigo17}. The adopted atmospheric parameters are given in Table \ref{tab:param}.

\subsection{Local thermodynamic equilibrium measurements}

\begin{table*}
\begin{center}   
\caption{Abundances of oxygen and of the light elements Li, Be, and B in our sample of stars for $\rm A(H)=12$.}
\label{tab:LiBeB}
\begin{tabular}{lcrcrrrrr}
\hline
\hbox{Star}    & [Fe/H]  & [O/H] & [O/H] &\hbox{A(Li)} &  A(Be)&  A(B)      & NLTE      & A(B)   \\
               &         & 777nm & OH    &             &       &  LTE       & corr      &  NLTE  \\
               &         &       &       &             &       &            & $S_{\rm H}=0.0$ & $S_{\rm H}=0.0$ \\
\noalign{\vskip 0.05cm}
\noalign{\hrule}
\noalign{\vskip 0.05cm}
\hbox{HD~19445}  & $-2.15$ & $-1.51$ & $-1.22$ & 2.29 & $-0.60$ & $+0.40$  & $+0.32$  & $+0.72$\\
\hbox{HD~76932}  & $-0.94$ & $-0.26$ & $-0.42$ & 2.12 & $+0.79$ & $+2.30$  & $+0.15$  & $+2.45$\\
\hbox{HD~84937}  & $-2.25$ & $-1.60$ & $-1.47$ & 2.22 & $-0.66$ & $< 0.00$ & $+0.22$  & $<+0.22$\\
\hbox{HD~94028}  & $-1.40$ & $-0.98$ & $-0.93$ & 2.30 & $+0.36$ & $+1.42$  & $+0.18$  & $+1.60$\\
\hbox{HD~140283} & $-2.57$ & $-1.67$ & $-1.81$ & 2.20 & $-1.01$ &$<$$-0.45$& $+0.23$  & $-0.22$\\
\hbox{HD~160617} & $-1.80$ & $-1.40$ & $-1.36$ & 2.32 & $-0.30$ &  $+0.55$ & $+0.22$  & $+0.77$\\
\hline	 
\end{tabular} 
\end{center}  
\end{table*}

With these parameters, atmospheric models were interpolated in the MARCS model grid by \citet{gustafsson08} and the abundances of the different elements were calculated \citep{peterson20,spite22,spite24}.
For all stars in Table \ref{tab:param}, synthetic spectra around the B lines were calculated using the code {\tt turbospectrum} from \citet{alvarez98} and \citet{plez12} and compared to the observed spectra. The results of these computations are given in Table \ref{tab:LiBeB} (Col. 7) together with the abundances of O, Li, and Be following \citet{spite24}.

\begin{figure*}
\begin{center}
\includegraphics [width=15.0cm]{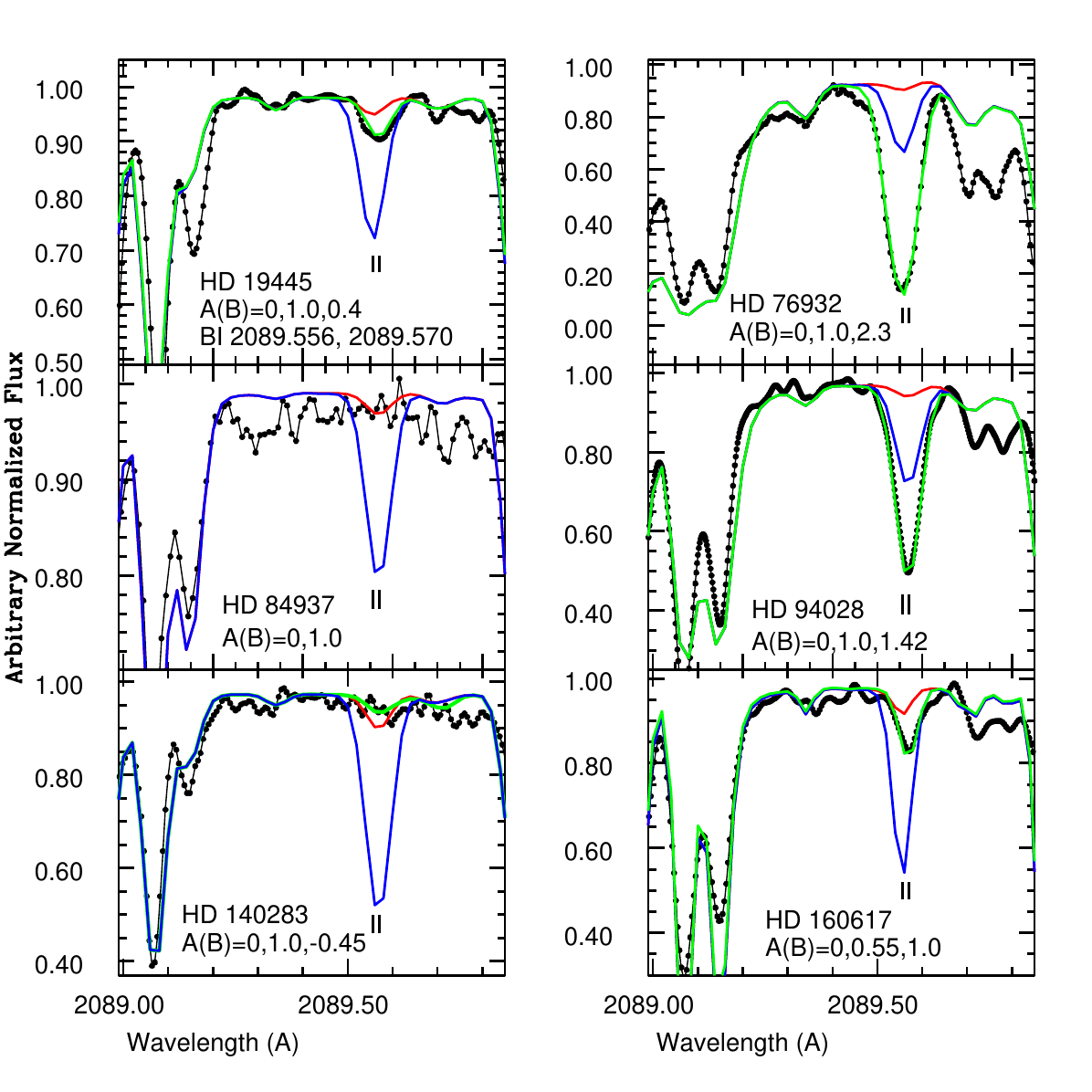}
\end{center}
  \caption{\ion{B}{I} 2089.556 and 2089.570\,{\AA} lines in the six stars in Table \ref{tab:param}.
  Observed spectra (black solid lines0, synthetic spectra computed with $\rm A(B)=0.0$ (red solid lines),
  +1.0 (blue solid lines), and final value (green solid lines).}
\label{fig:all-boro}
\end{figure*}

In Figure \ref{fig:all-boro} are shown the fits of the synthetic spectra to the clean \ion{B}{I} 2089.556 and 2089.570\,{\AA} lines in the six sample stars.

\subsection{Uncertainties}
In Table \ref{errors}, as an example, we have computed the uncertainties on the B abundance related to uncertainties in  stellar parameters, effective temperature
T$_{\rm eff}$ of  $\pm$100 K, gravity log \textit{g} of  $\pm$0.2\,dex, and 
microturbulence velocity v$_{\rm t}$ of $\pm$0.2 km s$^{-1}$. 
The resulting uncertainty on A(B) was computed  as the quadratic sum of these various sources of uncertainties; it is about 0.08\,dex and the error on [B/Fe] is negligible since a change in the model has about the same effect on the B and Fe abundances.\\ 
This error is about the same for all the stars in our sample since the parameters of the models of all our stars are very similar.\\
The errors on A(B) are mainly due to measurement errors, which are dominated by the uncertainty on the continuum placement. However, since the resolution of our spectra is high, this uncertainty is not very large (see upper panel in Fig.\,\ref{fig:hd94boro}). We adopted for all the stars the conservative values of a total error of 0.1\,dex on [Fe/H] and 0.2\,dex on A(B).

\begin{table}
  \caption{Boron abundance uncertainties for stellar parameters of $\Delta$T$_{\rm eff}$ = 100 K, $\Delta$log \textit{g} $=$ 0.2\,dex, and $\Delta$v$_{\rm t}$ = 0.2 km s$^{-1}$.  
} 
\label{errors}
\begin{flushleft}
\small
\tabcolsep 0.5cm
\begin{tabular}{lc@{}c@{}c@{}c@{}c@{}}
\noalign{\smallskip}
\hline
\noalign{\smallskip}
\hbox{star} & \hbox{$\Delta$T} & \hbox{$\Delta$log $g$} & 
\phantom{-}\hbox{$\Delta$v$_{t}$} & \phantom{-}\hbox{$\rm \Delta\, A(B)$} \\
\hbox{} & \hbox{100\,K} & \phantom{---}\hbox{0.2\,dex} & \phantom{---}\hbox{0.2 kms$^{-1}$} & \phantom{-}\hbox{($\sum$x$^{2}$)$^{1/2}$} & \\
\noalign{\smallskip}
\hline
\noalign{\smallskip}
HD~76932  & 0.07 &   0.00 & 0.04 & 0.08 \\
\noalign{\smallskip} 
\hline 
\end{tabular}
\end{flushleft}
 \end{table}

\subsection{NLTE computations}

The resonance lines of \ion{B}{I} are sensitive to non-local thermodynamic equilibrium (NLTE) effects.
Overionisation and overexcitation  are at play in the formation of the \ion{B}{I} resonance lines because, in particular, in our sample of metal-poor dwarfs, \ion{B}{I} is the minority ionisation stage. 
NLTE corrections of the 2089\,{\AA} line were calculated with the same hypotheses as in \citet{tan10}. 

Recently, \citet{voronov23} gave accurate rate coefficients for the \ion{B}{I} +  \ion{H}{I} collisions, but here, in a first approach, we have neglected these collisions (\ion{S}{H}=0). The NLTE corrections  are given  in Table \ref{tab:LiBeB} (Col. 8); they would decrease if the collisions of \ion{B}{I} with \ion{H}{I} were not negligible (\ion{S}{H}>0).

Furthermore the NLTE corrections are larger in hotter, lower gravity,  and more metal-poor stars.
In Table \ref{tab:LiBeB} we give  the NLTE value of the B abundance for \ion{S}{H}=0, in column 9.\\
Since in these stars, the Li abundance is constant and close to $\rm A(Li)=2.2$, we do not expect a stellar depletion of Be and B.

\section{Discussion}

\begin{figure*}
\begin{center}
    \resizebox{5.9cm}{!}      
    {\includegraphics [clip=true]{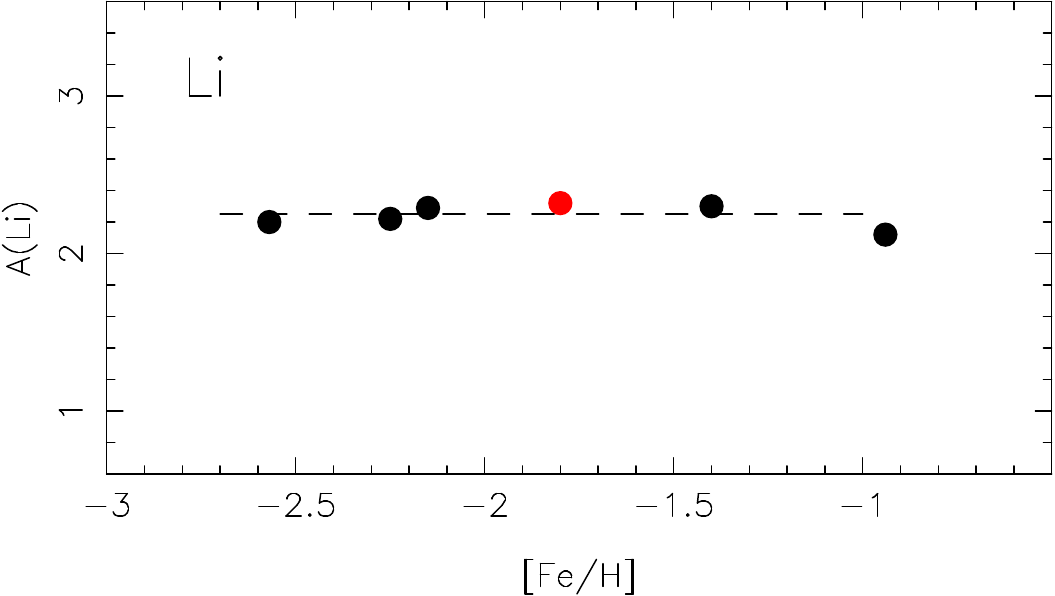}}
    \resizebox{5.9cm}{!}
    {\includegraphics [clip=true]{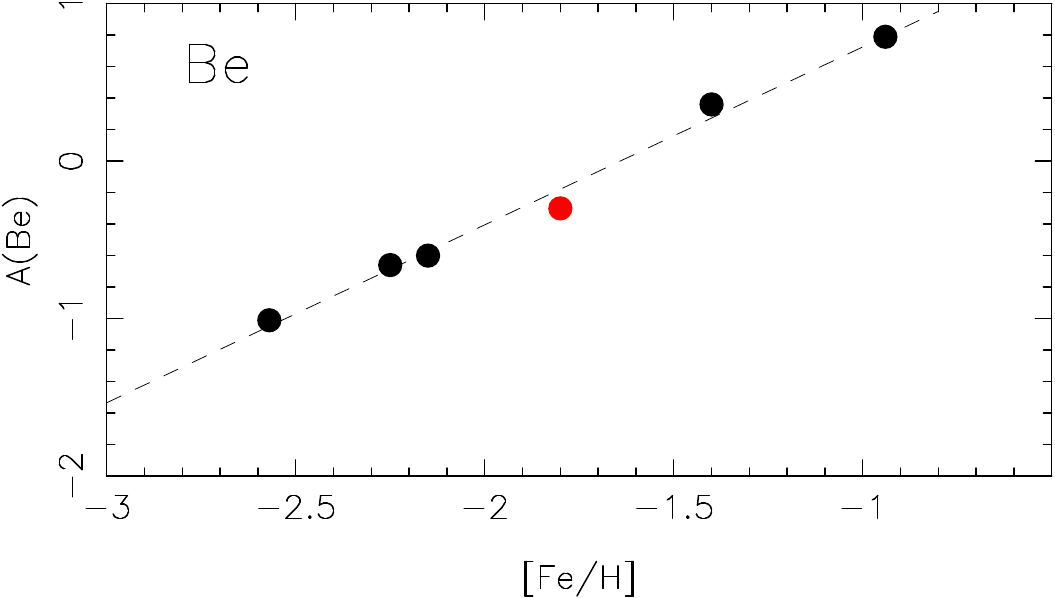}}
    \resizebox{5.9cm}{!}
    {\includegraphics [clip=true]{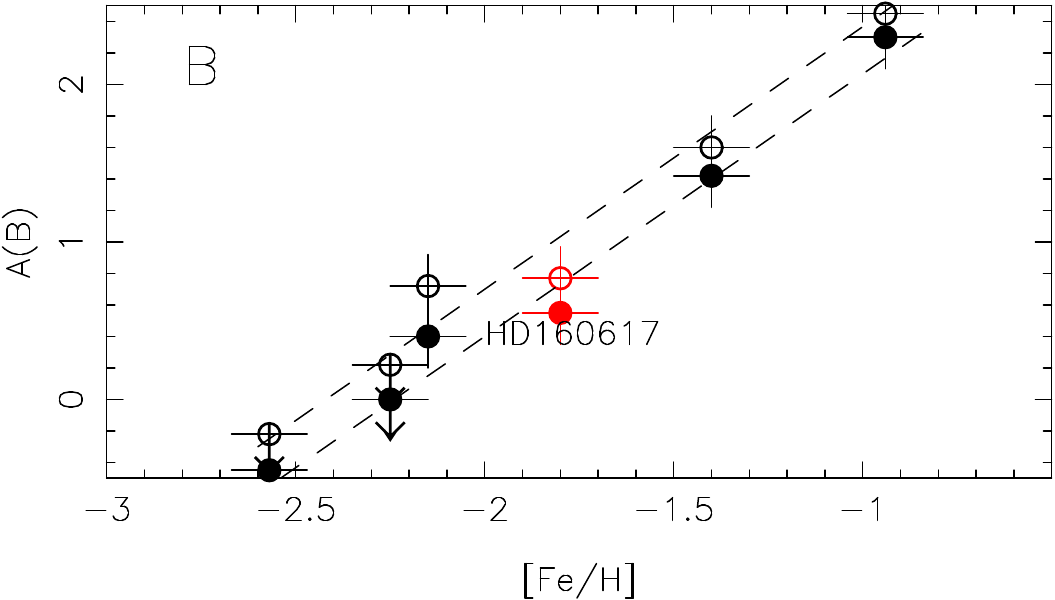}}
\end{center}
  \caption{A(Li), A(Be), and A(B) vs [Fe/H] in our sample of stars. The scale is the same in the three figures. In the right panel, for boron, the filled circles represent the LTE values of the B abundance and the open circles the NLTE values computed with \ion{S}{H}=0.0.  The red dot represents the N-rich star HD\,160617. The dashed lines are the linear regression lines.}
\label{fig:LiBeB}
\end{figure*}

\begin{figure}
    \resizebox{7.5cm}{!}
    {\includegraphics [clip=true]{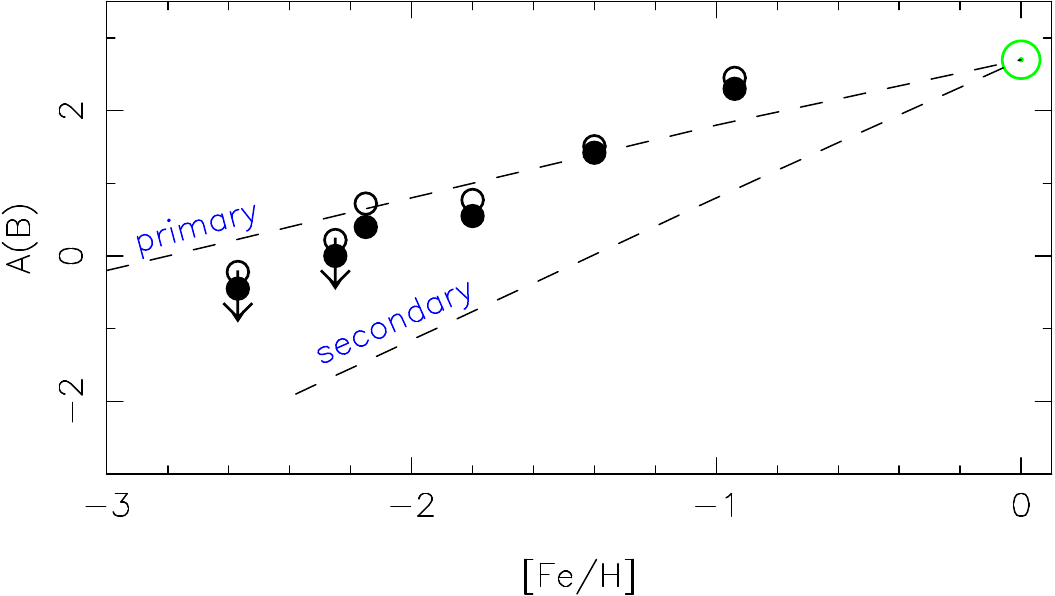}}
  \caption{Comparison of the relation log(B/H) vs [Fe/H] for the LTE and the NLTE values of the abundance of B (same symbols as in Fig. \ref{fig:LiBeB}). The dashed lines represent the predictions of  \citet{prantzos12} for a primary and a secondary production of B. The position of the Sun is indicated as a green symbol at $\rm [Fe/H]=0.0$  and A(B)=2.70
  \citep{lodders09,asplund21}. }
\label{fig:Prantzos}
\end{figure}

The evolution of Be abundance as a function of metallicity in our Galaxy suggests that Be is mainly produced as a primary element by energetic GCR in the environment of supernovae \citep{prantzos12}.
The formation of B is a little more complicated since there are two stable isotopes of B, $\rm ^{10}B$, and $\rm ^{11}B$. In meteorites $\rm ^{11}B / ^{10}B = 4$. Following \citet{prantzos12},  $\rm ^{10}B$ is almost completely produced by GCRs such as  Be, and 50\% of $\rm ^{11}B$ is also formed through bombarding by energetic GCR; $\rm ^{11}B$ can also be produced as primary by the $\nu$-process \citep[e.g.][]{olive94}.

\subsection{Behaviour of the abundance of boron versus [Fe/H]}
In Fig. \ref{fig:LiBeB} we plot the abundances A(Li), A(Be), and A(B) as a function of [Fe/H]. 
A(Be) and A(B) increase  with [Fe/H], but the slope of the relation A(B) versus [Fe/H] seems to be steeper than the slope of A(Be) versus [Fe/H]. The relations are as given in Eq. 1 below. The slope of A(B) versus [Fe/H] is significantly larger than 1:

\begin{equation} \label{eq1}
\begin{split}
     {\rm A(Be) = 1.129 [Fe/H] + 1.855 },\\
     {\rm A(B)~ =  1.644 [Fe/H] + 3.781 } .\\
\end{split}
\end{equation}

In Fig. \ref{fig:Prantzos} we compare our observations to the predictions of \citet{prantzos12} for a primary and a secondary formation of B with two hypotheses for the computation of the NLTE correction.
The mean value of A(B) in the interval $\rm -2.6<[Fe/H]<-1.0$ is in rather good agreement with the Prantzos's calculations\footnote{He computes $\rm log(B/H)=log(N_{B}/N_{H})=A(B)-12.0$} for a primary production of B and suggests that boron is mainly produced as primary such as Be by galactic cosmic rays (GCRs).
However, at very low metallicity ($\rm [Fe/H]<-2.5$) the abundance of B is lower than expected, and on the contrary at higher metallicity ($\rm [Fe/H]>-1.0$) it is higher than expected. 
If we extrapolate our measurements towards  the solar value we would expect $\rm A(Be)_{\odot}=1.85$ at $\rm [Fe/H]=0.0$, a value a little larger than the measured value $\rm A(Be)_{\odot}=1.38$ \citep{lodders09,asplund21}, but we would predict $\rm A(B)_{\odot}=3.80$, a value ten times larger than the value $\rm A(B)_{\odot}=2.70$ measured in the Sun following \citet{lodders09} and \citet{asplund21}.
\citet{garcia-lopez98}, who measured the abundance of B including solar metallicity stars, also suggested a change in slope at the halo-disk transition around $\rm [Fe/H]\sim-1.0$ \citep[see also][]{cunha00}.

\begin{figure}
\begin{center}
    \resizebox{7.5cm}{!}
    {\includegraphics [clip=true]{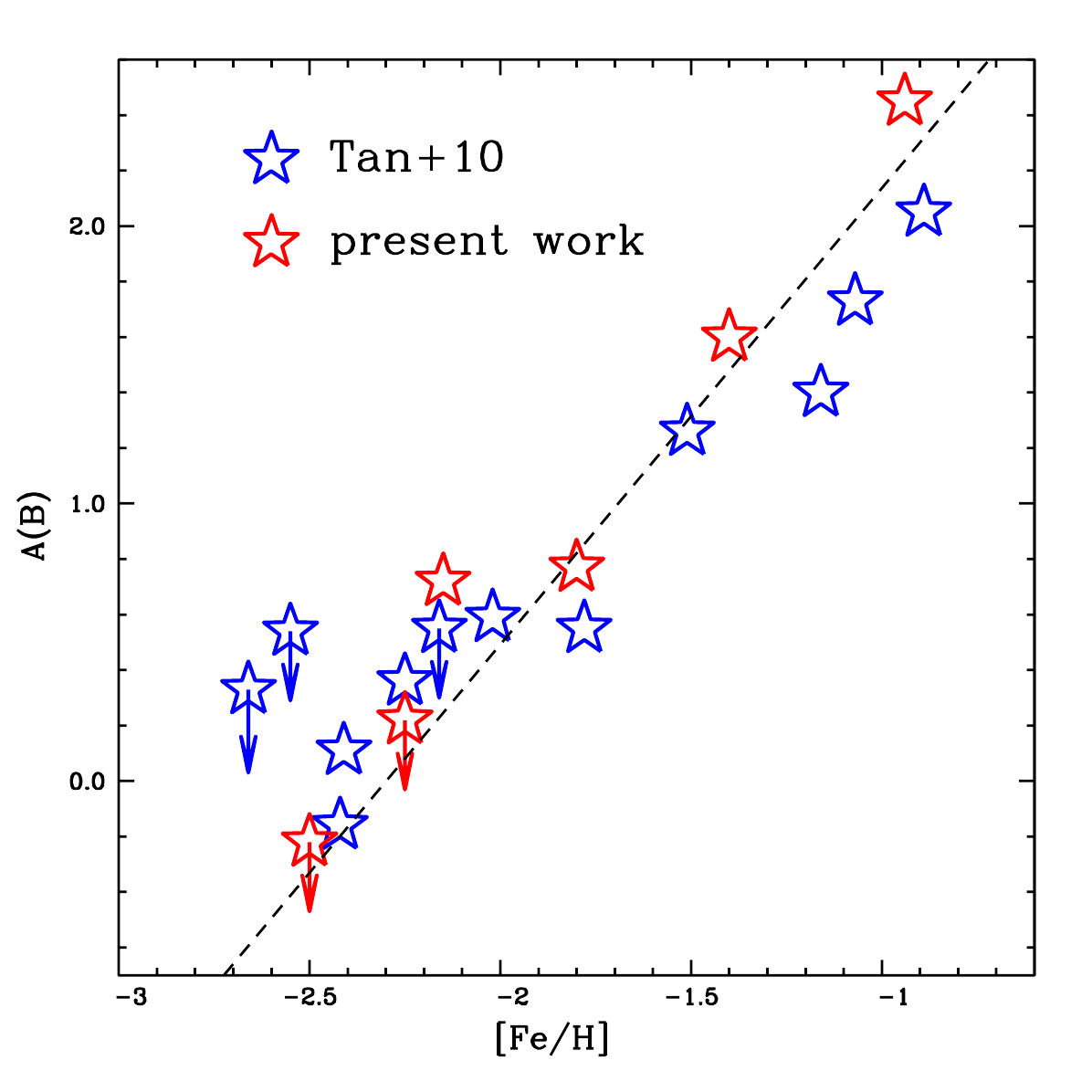}}
\end{center}
  \caption{A(B) vs [Fe/H] for our 6 stars (open red star markers) and
  12 Li-normal stars from \citet{tan10} (open blue \textbf{star markers)}.
  The dashed line corresponds to the value from Eq. 1 for boron.}
\label{fig:tan}
\end{figure}
 
\begin{figure}
\begin{center}
    \resizebox{7.0cm}{!}
    {\includegraphics [clip=true]{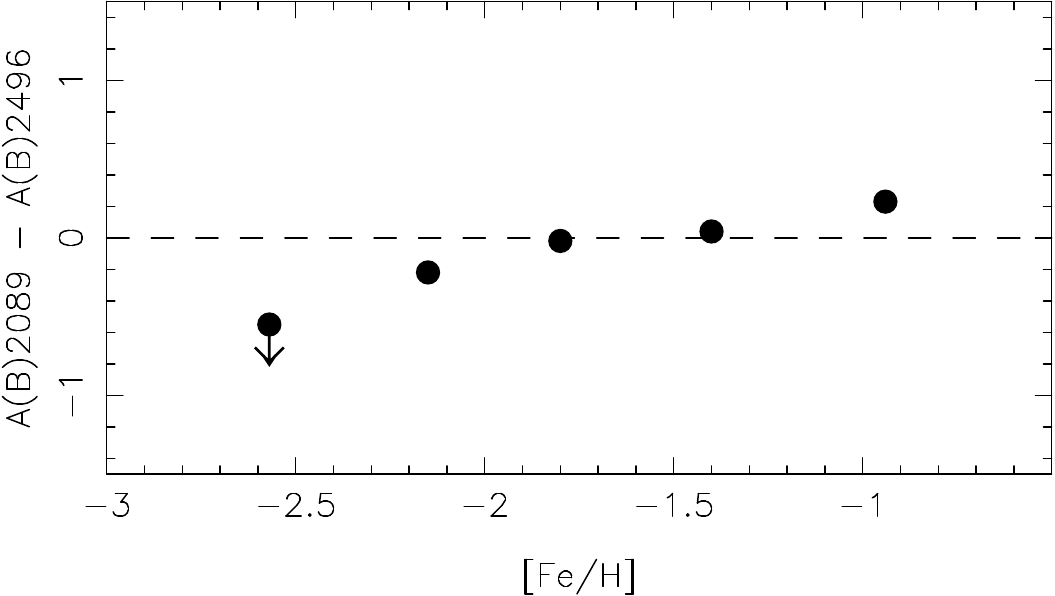}}
\end{center}
  \caption{Difference between the LTE B abundances from \citet{tan10} deduced from the 2496\,{\AA} line and scaled to our new stellar parameters and the abundances computed here (Table \ref{tab:LiBeB}) from the 2089\,{\AA} line.}
\label{fig:TanNou}
\end{figure}

In Fig. \ref{fig:tan} we  compare our measurements to the measurements of \citet{tan10} (based on the \ion{B}{I} 2496\,{\AA} line) for their set of metal-poor stars with a normal Li abundance (12 stars that include our six stars), the agreement is, a priori, rather good. We note that \citet{tan10} was able to measure the B abundance in only nine stars in their sample, and we were able to measure this abundance in  four stars out of six in ours;   for the remaining stars,  only an estimation of the upper limit was given. 
The precision of our measurements  is likely higher (see Fig. \ref{fig:hd94boro}), with little spread from the relation A(B) versus [Fe/H]. We see that the slope of our measurements appears to be steeper than for a pure primary production.\footnote{We did not compare our measurements to previous ones. All indeed \citep{duncan97,primas98,primas99,garcia-lopez98}, are based on the same spectra as used by \citet{tan10} and the analyses differ only in the stellar models (parameters) and the line lists used which were improved over time.}

\begin{figure}
\begin{center}
    \resizebox{7.0cm}{!}
    {\includegraphics [clip=true]{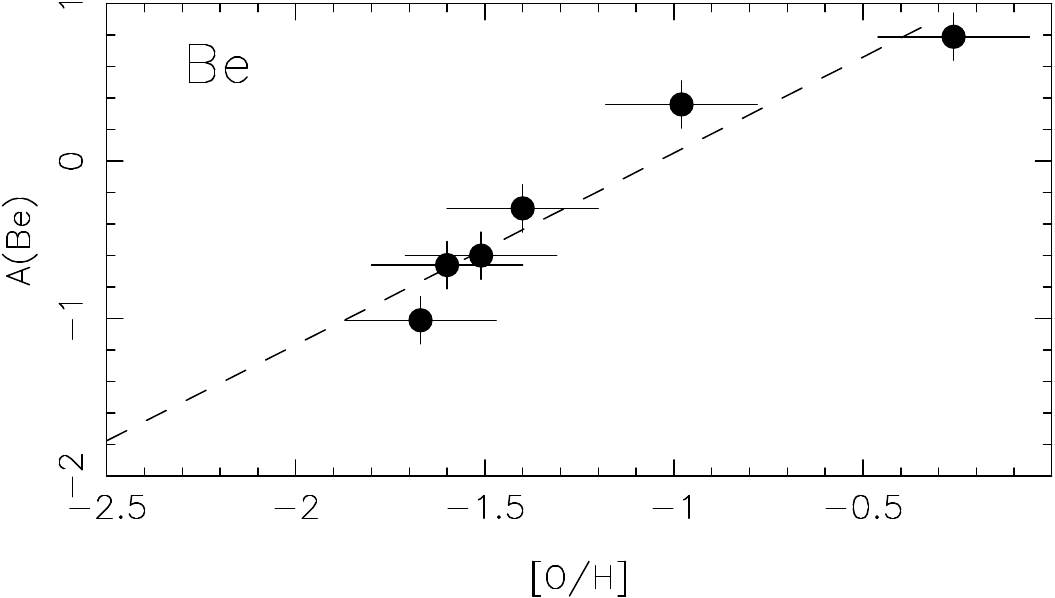}}
    \resizebox{7.0cm}{!}
    {\includegraphics [clip=true]{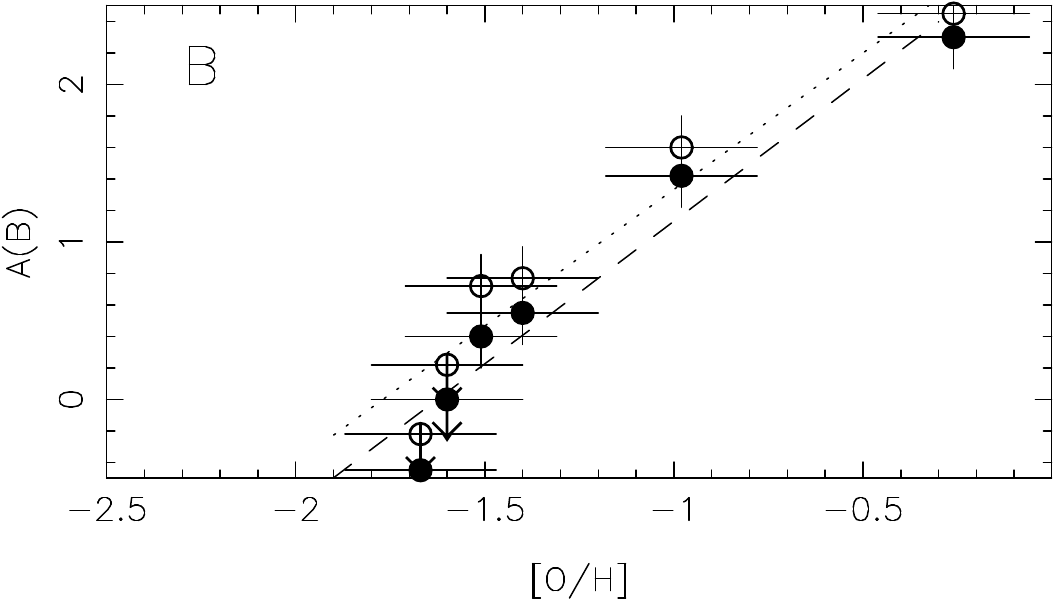}}
\end{center}
  \caption{Relation between the Be and B abundance with [O/H] deduced from the 777\,nm triplet. The symbols are the same as in Fig. \ref{fig:LiBeB}. In the top panel the dashed line represents the regression lines $\rm A(Be)=1.218[O/H]+1.269$. In the bottom panel the dashed line represents the regression line for the LTE computations $\rm A(B)=1.809[O/H]+2.940$ and the dotted line the regression line for the NLTE computations $\rm A(B)=1.732[O/H]+3.066$.}
\label{fig:BeB-O}
\end{figure}

Depending on the NLTE correction adopted for A(B), \citet{tan10} found a slope between  1.07 (NLTE values) and 1.17 (LTE values), and we found a slope between 1.58 (for NLTE values) and 1.65 (for LTE values).
We investigated the cause of this difference in slope between \citet{tan10} and our new measurements. If the B abundances from \citet{tan10} are scaled to our new stellar parameters, the differences in the B abundance 
for the five common stars shows a clear correlation with [Fe/H] (Fig. \ref{fig:TanNou}).\\ 
The differences in the NLTE abundance correction between \ion{B}{I}  2089 \AA~
and 2496 \AA~ does not exceed 0.14 dex for the sample stars. As can be seen in Figure 6, the difference in LTE B abundance between \ion{B}{I} 2089 \AA~ and 2496 \AA~ shows a clear
increasing trend with increasing [Fe/H] (from about --0.6\,dex at [Fe/H]=--2.57, to about +0.25\,dex at [Fe/H]=--0.94). 
The slope of B versus Fe relation based on \ion{B}{I} 2089 \AA~ is obviously higher than that based on B I 2496 A. The difference in the NLTE abundance correction between \ion{B}{I} 2089 \AA~ and 2496 \AA~ has only minor effects.\\
As a consequence the difference in the slope of \citet{tan10} and the present paper is clearly the result of the use of the \ion{B}{I} 2089\,{\AA} line instead of the 2496\,{\AA} line and not of the slight difference in the model parameters or the NLTE corrections. 

\citet{primas10} and \citet{tan11} found, around $\rm [Fe/H]=-1$, systematic differences in the Be abundances among the $\rm high-[\alpha]$ and the $\rm low-[\alpha]$ populations. We could expect similar differences for B. It is thus important to note that all the metal-poor stars  studied here belong to the $\rm high-[\alpha]$ population.

\begin{figure}
\begin{center}
    \resizebox{7.0cm}{!}
    {\includegraphics [clip=true]{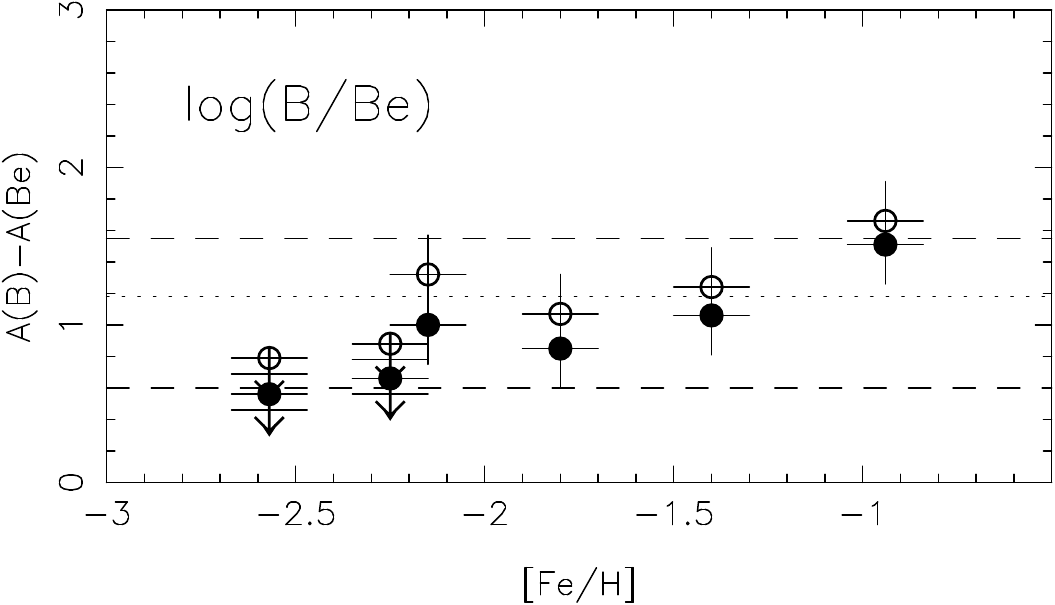}}
\end{center}
\caption{Relation between log(B/Be) and the abundance of Fe. The symbols are the same as in Fig. \ref{fig:LiBeB}. The dotted line represents the mean value of log(B/Be) found by \citet{tan10} 
($\rm log(B/Be) \approx 15$). The two dashed lines represent the limits for our sample of stars that correspond to $\rm 4<log(B/Be)<35$.}
\label{fig:BeBratio}
\end{figure}

\subsection{The case of HD\,160617}
In Fig. \ref{fig:LiBeB}, the red dot represents HD\,160617, a star known to be nitrogen-rich \citep{laird85,spite22}. This star has a normal abundance of Li and Be; however,   from the \ion{B}{I} 2496\,{\AA} line, \citet{primas98} found it to be depleted in B by about 0.5\,dex. This anomaly had presented a real challenge to theoreticians at the time. However, \citet{tan09,tan10} found that in this star, considering the uncertainties, the Be abundance and  the B abundance were in reasonable agreement with the general trend.
From our new spectra, the boron abundance appears also to be normal in this star, and so this discrepancy is now completely solved. 

\subsection{Behaviour of the abundance of boron versus [O/H]}
Since B and Be are supposed to be formed by spallation of CNO nuclei we can expect a very tight relation between the abundance of B, Be, and O. Unfortunately, very precise measurements of the oxygen abundance are often difficult to obtain in warm metal-poor dwarfs and they often differ from the 
oxygen lines used \citep[see e.g.][]{tan10}. 
In Table \ref{tab:LiBeB} Cols. 3 and 4, we give the abundance of oxygen deduced from the \ion{O}{I} 7774\,{\AA} triplet and from the ultraviolet OH lines following \citet{spite22}.

In Fig. \ref{fig:BeB-O} we  plot A(Be) and A(B) (LTE and NLTE values) versus [O/H] in place of [Fe/H]. The O abundance in this figure is deduced from the 777\,nm triplet, as was done by \citet{tan10}. Here again the scatter is small, but the slope of A(B) versus [O/H] is steeper than in \citet{tan10}.
Depending on the NLTE correction adopted, the slope varies between 1.73 (NLTE) and 1.81 (LTE) in place of 1.36 and 1.44 in \citet{tan10}.

\subsection{The B-to-Be ratio}
Finally, in Fig. \ref{fig:BeBratio} we  plot $\rm log(B/Be)=A(B)-A(Be)$ as a function of [Fe/H].
Since at low metallicity the enrichment of B deviates more from a mean primary production than that of Be, the ratio B/Be seems to increase for increasing [Fe/H] and [O/H]. It is only about 4 at $\rm [Fe/H]=-2.5$, but 35 at $\rm [Fe/H]=-1$.
The dotted line in this figure represents the mean value of this ratio found by \citet{tan10} for the nine stars in their sample of metal-poor stars, and the two dashed lines indicate the limits of the B/Be value for our sample of stars.

\section{Conclusions}

The small number of stars analysed (six) is here compensated by a much higher
precision relative to previous derivations of boron abundances.
The behaviour of the boron abundance at low metallicity was always based on a small number of stars (about ten), and it is the first time in 25 years that new HST spectra have been analysed. 

Until the early 2000s the debate on Be and B
was on the primary nature of the A(Be) and A(B) versus [Fe/H] relations.
The question was, given that the Be and B production
from spallation in the ISM should depend on the abundance of
the C, N, and O nuclei, that a secondary behaviour should be expected.
The discussions were then on the [Fe/H] versus [O/H] relation
\citep[e.g.][]{fieldsolive99,king01}.
This issue was essentially solved with the work by Prantzos (2012),
by coupling wind yields of the
Geneva models of rotating mass-losing stars with a detailed galactic
chemical evolution code. He was able to reproduce the
A(Be) versus [Fe/H] relation, and 70\% of A(B) versus [Fe/H], which is complemented
with $\nu$-process producing the remaining 30\% of $^{11}$B.
Based on the previous observations, he confirmed the primary behaviour of A(Be) versus [Fe/H], and
A(B) versus [Fe/H].

In the present work, we presented a step further, given the high precision of our B abundances. We see that 
in the interval $\rm -2.6<[Fe/H]<-1.0$  the mean value of the B abundance is compatible with the Prantzos model based on a primary production of B. 
However, at very low metallicity ($\rm [Fe/H]<-2.5$) the abundance of B is significantly lower than that expected by the Prantzos's model. 
It seems that in the interval $\rm -2.6<[Fe/H]<-1.0$ the B enrichment becomes very efficient, and at $\rm [Fe/H]=-1.0$ the B abundance is already close to the solar abundance.
Another consequence is that as soon as $\rm [Fe/H]>-1.0$ the B enrichment should be less efficient. Between $\rm [Fe/H]=-2.5$ and $-1.5$ the abundance of B increases by more than 1.5\,dex, and between $\rm [Fe/H]=-1.0$ and the Sun ($\rm [Fe/H]=0.0$) the abundance of B  increases only  by less than 0.4\,dex. The B enrichment seems to present a break at about $\rm [Fe/H]=-1$. This behaviour had been already suggested \citep[see][]{garcia-lopez98,cunha00}, but with a higher precision of the B abundance this unexpected effect seems to be more firmly established.

\citet{prantzos12} galactic chemical evolution models suggest that 70\% of $^{11}$B is due to cosmic ray spallation, and the
remaining 30\% are due to the $\nu$-process. \citet{kusakabe19} find a metallicity-dependent production of light elements from the $\nu$-process.
The relation A(B) versus metallicity found here might be accommodated with a initial production of
B from the $\nu$-process in massive stars at the early stages of the Galaxy life \citep{woosley90,woosley95,kusakabe19}, 
and supernovae type Ic at their lower mass end as computed by \citet{nakamura10,nakamura10b},
provided that this enrichment in $^{11}$B is metallicity dependent, decreasing with time.

Further theoretical investigations and new observations to confirm,  specify, and explain the A(B) versus metallicity relation found in the present work, would be of great interest. 
Currently, new observations with HST/STIS for more stars (fainter than the stars observed) would require very large exposure times, and thus it would be extremely difficult to obtain spectra of a larger number of metal-poor dwarfs.
In a near future this will become possible with the next-generation flagship space-based  Habitable World Observatory (HWO) 
telescope equipped with the spectropolarimeter POLLUX \citep{muslimov24}, for example.

\begin{acknowledgements}
We greatly thank Ruth Peterson for the collaboration leading to the acquisition of the
high-quality data used in this work, and for the data reduction.
We also thank Nicolas Prantzos for useful discussions.
BB acknowledges partial financial support from the agencies FAPESP, CNPq
and CAPES - financial code 001.
KT is supported by the National Key R\&D Program of China (No.~2024YFA1611903) and the Strategic Priority Research Program of Chinese Academy of Sciences (No.~XDB1160103).
This research is based on observations made with the NASA/ESA Hubble Space Telescope obtained from the Space Telescope Science Institute, which is operated by the Association of Universities for Research in Astronomy, Inc., under NASA contract NAS 5–26555. These observations are associated with program(s) 
7348 (PI: Edvardsson), 8197 and 9804 (PI: Duncan), 14161 (PI: Peterson), 14672 (PI: Peterson). This work has made use of the VALD database, operated at Uppsala University, the Institute of Astronomy RAS in Moscow, and the University of Vienna.

\end{acknowledgements}

\bibliographystyle{aa} 
\bibliography{biblioboron}

\end{document}